\documentclass[aps, prb, reprint, longbibliography, superscriptaddress, floatfix]{revtex4-2}
\usepackage{graphicx}% Include figure files
\usepackage{dcolumn}% Align table columns on decimal point
\usepackage{bm}% bold math
\usepackage{physics}
\usepackage{amsfonts}
\usepackage{mathrsfs}
\usepackage{subfigure}
\usepackage[svgnames]{xcolor}
\usepackage[colorlinks=true,linkcolor=NavyBlue,anchorcolor=red,citecolor=NavyBlue, urlcolor=NavyBlue]{hyperref}
\usepackage{color}
\usepackage{xcolor}
\DeclareMathAlphabet{\pazocal}{OMS}{zplm}{m}{n}
\usepackage{amsmath}
\usepackage{amssymb}
\usepackage[T1]{fontenc}
\usepackage{makecell}

\begin{document}

\title{Electron and phonon spectrum in a metallic nanohybrid}

\author{Debraj Bose}
\affiliation{Harish-Chandra Research Institute (A CI of Homi 
Bhabha National Institute), Chhatnag Road, Jhusi, Allahabad 211019}

\author{Saheli Sarkar}
\affiliation{Harish-Chandra Research Institute (A CI of Homi 
Bhabha National Institute), Chhatnag Road, Jhusi, Allahabad 211019}

\author{Pinaki Majumdar}
\affiliation{School of Arts and Sciences, Ahmedabad 
University, Ahmedabad, India 380009}

\pacs{75.47.Lx}
\date{\today}

\begin{abstract}
Recent experiments on metallic nanohybrids have revealed unusually strong
electron–phonon effects emerging from nanoscale interfaces, despite the
weak coupling character of the constituent bulk materials. Motivated by
these observations, we investigate the electronic and lattice spectral
properties of an inhomogeneous electron–phonon system in which strong
coupling is confined to interfacial regions embedded in a weakly coupled
metallic background.
Using a real-space formulation of the Holstein model combined with
Langevin dynamics for lattice equilibration, we compute both electronic
and phonon spectral functions in the presence of spatially varying
coupling. We find that increasing the fraction of interfacial sites
leads to a pronounced broadening of electronic spectral features,
reflecting enhanced quasiparticle scattering from lattice distortions,
but leaves the underlying band dispersion largely intact.
Simultaneously, the phonon spectrum exhibits significant softening and
damping, originating from strongly distorted interfacial regions.
These modifications result in a redistribution of the Eliashberg
spectral function toward low frequencies, producing a substantial
enhancement of the effective electron–phonon coupling constant.
Our results demonstrate that spatial inhomogeneity alone can strongly
renormalize both electronic and lattice spectra, and provide a
microscopic framework for understanding interface-driven 
transport and interaction effects in metallic nanohybrids.
\end{abstract}

\maketitle

\section{Introduction}

Metallic systems composed of nanoscale hybrid \cite{arind1} structures 
can exhibit
electronic properties that differ qualitatively from those of their
constituent bulk materials. In particular, recent experiments on
Ag–Au nanohybrids \cite{arind2,arind3} have reported unusually large 
temperature-dependent
resistivity and signatures of enhanced electron–phonon (EP) coupling,
despite both Ag and Au being weak-coupling metals in the bulk.
Interfaces can act as active regions where interactions are strongly 
modified, generating spatially inhomogeneous electronic 
environments that feed back onto lattice dynamics.

A central question raised by these experiments is how spatially
localized enhancements of electron–phonon coupling influence the
electronic and lattice dynamics \cite{ziman1960,allen2006,allen2000}
 of an otherwise weakly interacting
metal. In particular, it is not obvious whether such inhomogeneity
primarily renormalizes the electronic structure, modifies quasiparticle
lifetimes, or leads to qualitative changes in lattice excitations.
Standard effective-medium descriptions, which average over microscopic
details, are not suited to capture the interplay between strong local
coupling and extended electronic states.

An important aspect of this problem, which we emphasize in this 
work, is that lattice excitations provide a particularly sensitive 
probe of spatially inhomogeneous electronic environments. 
In a homogeneous system, phonon renormalization can be 
understood in terms of momentum-resolved screening 
processes. In contrast, in an inhomogeneous system the relevant 
quantity is a spatially varying electronic susceptibility, which 
leads to a distribution of local lattice stiffness and damping. As
 a result, phonons no longer correspond to well-defined normal 
 modes, but instead reflect fluctuations in a self-generated, 
 spatially structured energy landscape. Understanding how 
 such a landscape modifies phonon spectra is therefore 
 central to describing the interplay of disorder and 
 interaction in nanohybrid systems.

In this work we address this problem using an inhomogeneous
electron–phonon model in which strong coupling is confined to
interfacial regions embedded within a weakly coupled background.
We employ a real-space formulation of the Holstein model
 \cite{bose2024}, combined
with Langevin dynamics to obtain equilibrium lattice configurations,
and compute both electronic and phonon spectral functions from the
resulting configurations.

{It is useful to distinguish this inhomogeneous Holstein
 framework from the disordered Holstein model
  \cite{sergeev2000electron,bronold2001dynamics,
 capone2003,xiao2021,ciuchi1997,di2014strong,stolpp2020charge}.
  In the latter, disorder typically enters as random variations in 
  on-site energies or coupling strengths distributed throughout
   the system, leading to spatially uncorrelated scattering and 
   possible localization effects \cite{tozer2014localization}. In
    contrast, the inhomogeneous model considered here involves
     a structured spatial modulation, where strong electron–phonon
      coupling is confined to specific interfacial regions. As a result,
       the physics is governed not by randomness alone but by 
       the coexistence of strongly and weakly coupled domains, 
       enabling coherent electronic motion across the system
        while experiencing enhanced scattering at well-defined interfaces.}

We show that spatial inhomogeneity produces a distinct spectral
signature: the electronic dispersion remains largely intact, but
quasiparticle peaks broaden significantly due to scattering from
interface-induced lattice distortions. At the same time, phonon modes
exhibit pronounced softening and damping, leading to enhanced low-frequency
weight in the Eliashberg spectral function 
\cite{aperis2018,chubukov2020} and a substantial increase in
the effective electron–phonon coupling \cite{mandal2025}. 
More broadly, our work highlights a regime in which interactions are
not uniformly distributed but are instead concentrated in spatially
localized regions, leading to emergent behavior that cannot be captured
within homogeneous or effective-medium frameworks.

To summarize our main qualitative findings, we identify three key
 effects arising from interfacial inhomogeneity. (i) The electronic
  spectral function exhibits a pronounced broadening of 
  quasiparticle peaks with increasing interfacial fraction, while 
  the underlying band dispersion remains largely unchanged,
   indicating enhanced scattering without significant band 
   reconstruction. (ii) The phonon spectrum exhibits strong 
   softening and damping, reflecting fluctuations in a spatially
    inhomogeneous lattice environment generated by the
     electronic degrees of freedom. In particular, phonon 
     excitations probe a distribution of local stiffness and 
     decay channels associated with interfacial regions.
      (iii) These changes lead to a redistribution of the 
      Eliashberg spectral function toward low frequencies 
      and a substantial enhancement of the effective
       coupling constant $\lambda$.

This paper is organized as follows. Sec.~II introduces
 the model and method, followed by electronic spectral 
 properties in Sec.~III. The phonon response is discussed 
 in Sec.~IV, and the Eliashberg function is analyzed in 
 Sec.~V. We then outline broader implications of our
  results in Sec.~VI, before summarizing in Sec.~VII.

\section{Model and Computational Method}

\subsection{Model}

To describe the Au–Ag nanohybrid system we 
use an inhomogeneous
Holstein model defined on a two–dimensional
 square lattice \cite{bose2024}.
The total Hamiltonian is written as
\begin{equation}
H = H_{\rm el} + H_{\rm ph} + H_{\rm el-ph},
\end{equation}
where the three terms respectively represent the
 Hamiltonian for the electronic kinetic energy,
the Hamiltonian for the lattice degrees of freedom,
 and the Hamiltonian for the electron–phonon interaction.
The electronic part is
\begin{equation}
H_{\rm el} =
- t \sum_{\langle ij \rangle} \left(c^{\dagger}_{i}c_{j} + h.c.\right)
- \mu \sum_{i} n_i ,
\end{equation}
where $c^{\dagger}_i$ ($c_i$) creates 
(annihilates) an electron at site $i$,
$n_i = c^{\dagger}_i c_i$ is the local electron density,
$t$ is the nearest–neighbor hopping amplitude,
and $\mu$ controls the electron filling.

The lattice degrees of freedom are described
 by classical phonon
coordinates $X_i$ and conjugate momenta $P_i$,
\begin{equation}
H_{\rm ph} =
\sum_i
\left(
\frac{P_i^2}{2M}
+
\frac{K}{2}X_i^2
\right),
\end{equation}
where $M$ is the ionic mass and $K$ is the lattice stiffness.
The bare phonon frequency is $\omega_0 = \sqrt{K/M}$.

Electron–phonon coupling is incorporated through the Holstein
interaction \cite{jansen2023,ying2024,nosarzewski2021}
\begin{equation}
H_{\rm el-ph} = - \sum_i g_i n_i X_i ,
\end{equation}
where $g_i$ is the local electron–phonon coupling strength.

Spatial inhomogeneity is introduced 
\cite{xiao2021,emin1994,kumar2005} through
 the distribution of $g_i$.
Sites belonging to the bulk regions are 
assigned a weak coupling $g_1$,
while sites located at the Au–Ag interface are 
assigned a stronger
coupling $g_2$.
The fraction and spatial arrangement of interfacial
sites is controlled by a parameter $f$, which is the
concentration of Ag sites embedded in the Au matrix.

Throughout this work we set the hopping
 amplitude $t=1$ as the unit
of energy. Unless otherwise stated we use parameters
$g_1 = 0.2$, $g_2 = 1.6$, and $\omega_0 = 0.2$,
which represent weak electron–phonon coupling in the bulk and
enhanced coupling at the interface.

\subsection{Generation of equilibrium lattice configurations}

Equilibrium lattice configurations at temperature $T$ are obtained
by evolving the lattice degrees of freedom using Langevin dynamics
\cite{lang-lu,bhattacharyya2019,bhattacharyya2020}.
The equation of motion for the lattice displacement $X_i(t)$ is
\begin{equation}
M\frac{d^2 X_i}{dt^2}
=
- K X_i
-
\frac{\partial \langle H_{\rm el} \rangle}{\partial X_i}
-
M\gamma \frac{dX_i}{dt}
+
\xi_i(t),
\end{equation}
In the equation above $t$ is time 
and is not to be confused 
with the hopping parameter.  
$\gamma$ is a damping coefficient 
and $\xi_i(t)$ is a stochastic
noise term.
The electronic force acting on the 
lattice is determined from the
instantaneous electronic density,
\begin{equation}
\frac{\partial \langle H_{\rm el} \rangle}{\partial X_i}
=
- g_i \langle n_i \rangle .
\end{equation}
The local density is computed from the instantaneous eigenstates of
the electronic Hamiltonian,
\begin{equation}
\langle n_i \rangle =
\sum_m |\psi_{im}|^2 f(\epsilon_m),
\end{equation}
where $\epsilon_m$ and $\psi_{im}$ are
 the eigenvalues and eigenvectors
of the electronic Hamiltonian for a given
 lattice configuration and
$f(\epsilon)$ is the Fermi–Dirac distribution.
The noise satisfies the 
fluctuation–dissipation relation
\begin{equation}
\langle \xi_i(t) \rangle = 0,
\qquad
\langle \xi_i(t)\xi_j(t') \rangle =
2\gamma k_B T \delta_{ij}\delta(t-t').
\end{equation}
The Langevin dynamics (LD) scheme 
employed here operates in the 
adiabatic limit, where the characteristic phonon frequency is 
small compared to the electronic energy scales, i.e., $\omega_0 \ll t$, 
consistent with earlier formulations of the 
LD approach~\cite{bhattacharyya2019}.

% --------------------------------------------------------
\begin{figure*}[t]
\centerline{
\includegraphics[width=15cm,height=4.5cm]{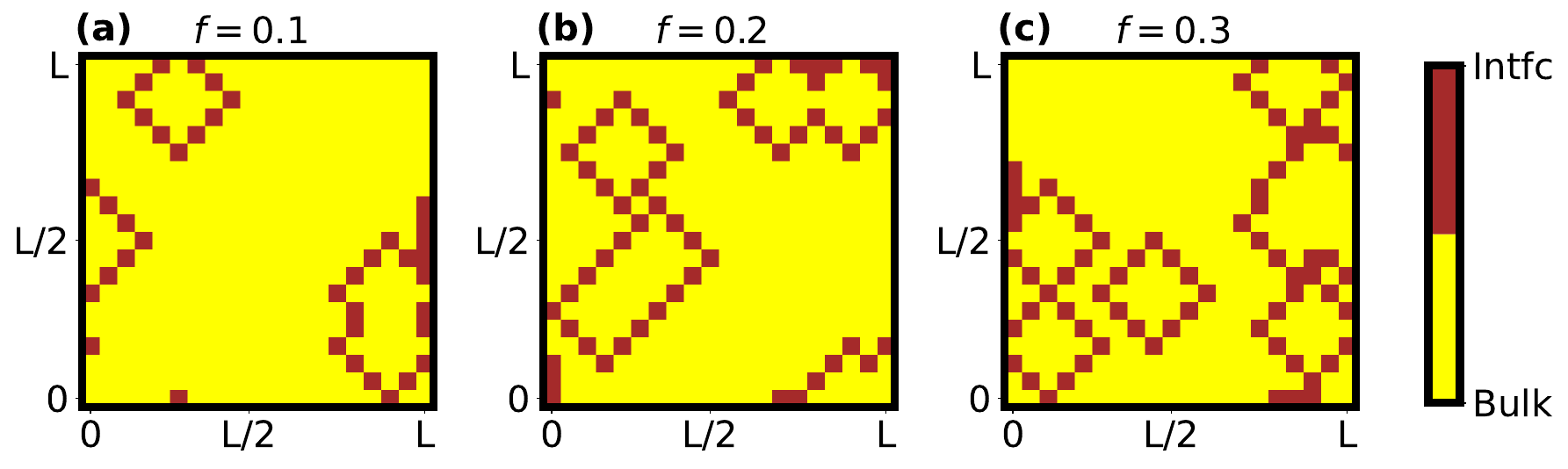}
}
\caption{Spatial structure of the inhomogeneous electron–phonon
model for interfacial fractions $f = 0.1, 0.2,$ and $0.3$. Boundary
sites (red) denote interfacial regions with enhanced coupling
$g = g_2$, while bulk sites (yellow) have weak coupling $g = g_1$.
Increasing $f$ increases both the density and connectivity of
strongly coupled regions, thereby tuning the degree of spatial
inhomogeneity in the system.}
\label{Fig:SpatialStructure}
\end{figure*}
% --------------------------------------------------------

Although the equations of motion are formulated at 
finite temperature, the spectral results correspond to 
the $T \to 0$ limit. The system is annealed from high
 temperature down to $T = 5 \times 10^{-4} t$, 
 ensuring well-equilibrated configurations.

\section{Electronic Spectra: Lifetime Broadening without Band Reconstruction}

We begin by describing the spatial
structure of the system on which the
electronic spectra are calculated. 
Fig.\ref{Fig:SpatialStructure} shows representative
configurations of the interface for three values of the parameter
$f=0.1$, $0.2$, and $0.3$. The boundary sites (colored brown) 
denote the
interfacial regions where the electron--phonon coupling is enhanced
($g=g_2$), while the bulk sites (yellow) have the weaker coupling
$g=g_1$. As $f$ increases the number and connectivity of interfacial
regions grow, leading to a progressively more
 inhomogeneous electronic
environment.

Having established the underlying spatial structure, we now describe how
the electronic spectral properties are computed. All spectral quantities
reported in this section correspond to equilibrium lattice configurations
in the low-temperature limit $T \rightarrow 0$.

\subsection{Calculation of $G(\mathbf{k},\omega)$ and $A(\mathbf{k},\omega)$}

For a given cluster geometry the $\{X_i\}$ are obtained from
the Langevin dynamics described in Sec.~II. The electronic Hamiltonian
is diagonalized exactly to obtain the eigenvalues $E_n$ 
and the eigenvectors
$\psi_i^{(n)}$.
The retarded real-space Green's function is constructed as
\begin{equation}
G^R_{ij}(\omega)
=
\sum_n
\frac{\psi_i^{(n)} \psi_j^{(n)*}}
{\omega - E_n + i\eta},
\end{equation}
where $\eta$ is a small Lorentzian broadening 
parameter, and $\eta=0.001$.
To obtain momentum-resolved information we 
perform a Fourier transform
of the real-space Green's function,
\begin{equation}
G^R(\mathbf{k},\omega)
=
\sum_{ij}
e^{-i\mathbf{k}\cdot(\mathbf{r}_i-\mathbf{r}_j)}
G^R_{ij}(\omega).
\end{equation}
% --------------------------------------------------------
\begin{figure}[b]
\centering
\includegraphics[width=8.4cm,height=7.2cm]{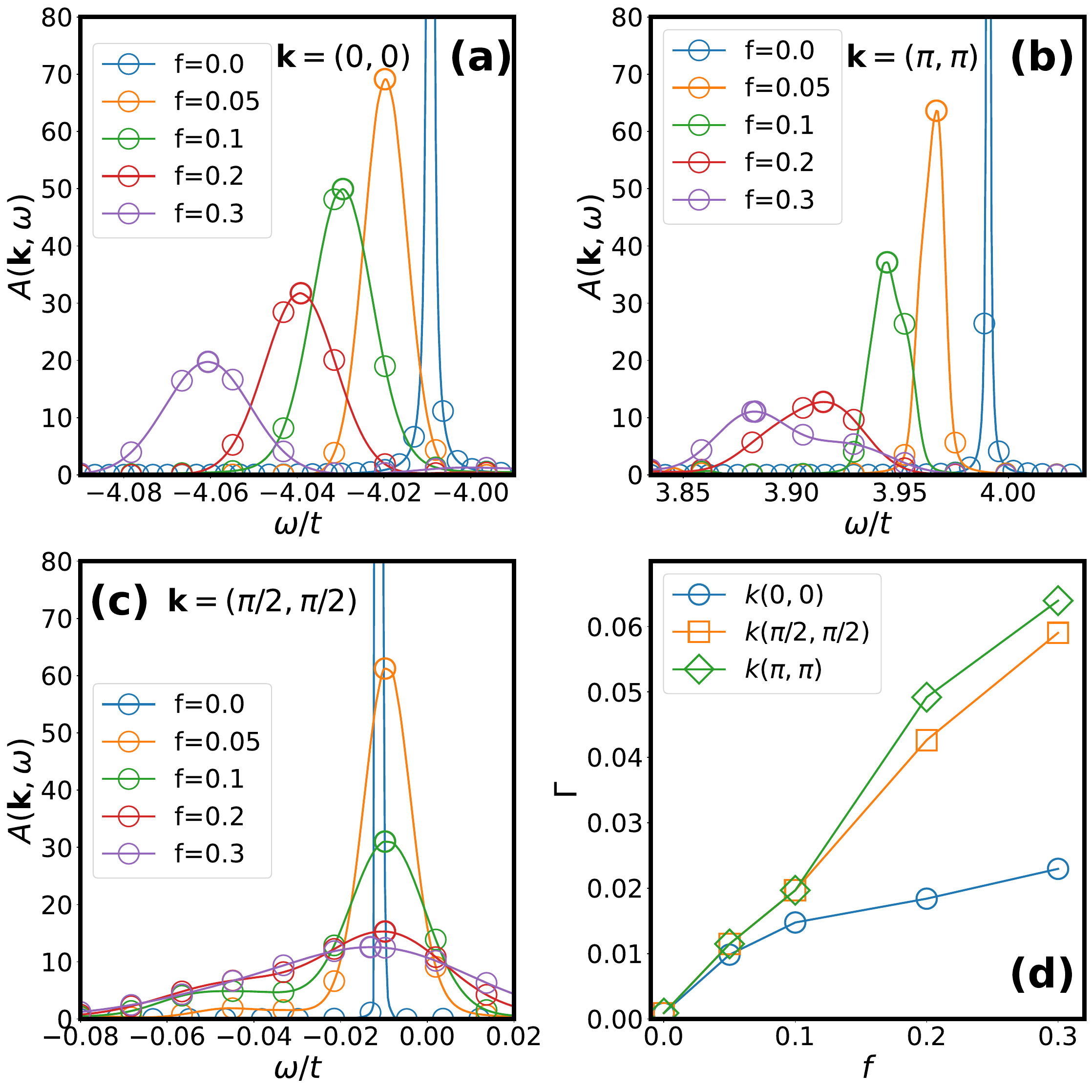}
\caption{Momentum-resolved electronic spectral function
 $A(\mathbf{k},\omega)$ at representative momenta: (a) $k=(0,0)$,
 (b) $(\pi,\pi)$, and (c) $(\pi/2,\pi/2)$, for increasing interfacial
  fraction $f$. Quasiparticle peaks broaden systematically
  with increasing $f$, indicating enhanced scattering from
  interface-induced lattice distortions. (d) Extracted linewidth
  $\Gamma$ as a function of $f$ shows a monotonic increase
  for all momenta. Notably, the peak positions shift only weakly,
  demonstrating that interfacial inhomogeneity primarily
  reduces quasiparticle lifetimes without significantly
  renormalizing the underlying band structure.}
  \label{Fig:electronspectra_momenta}
\end{figure}
% --------------------------------------------------------
% --------------------------------------------------------
\begin{figure}[t]
\centering
\includegraphics[width=6.6cm,height=5.2cm]{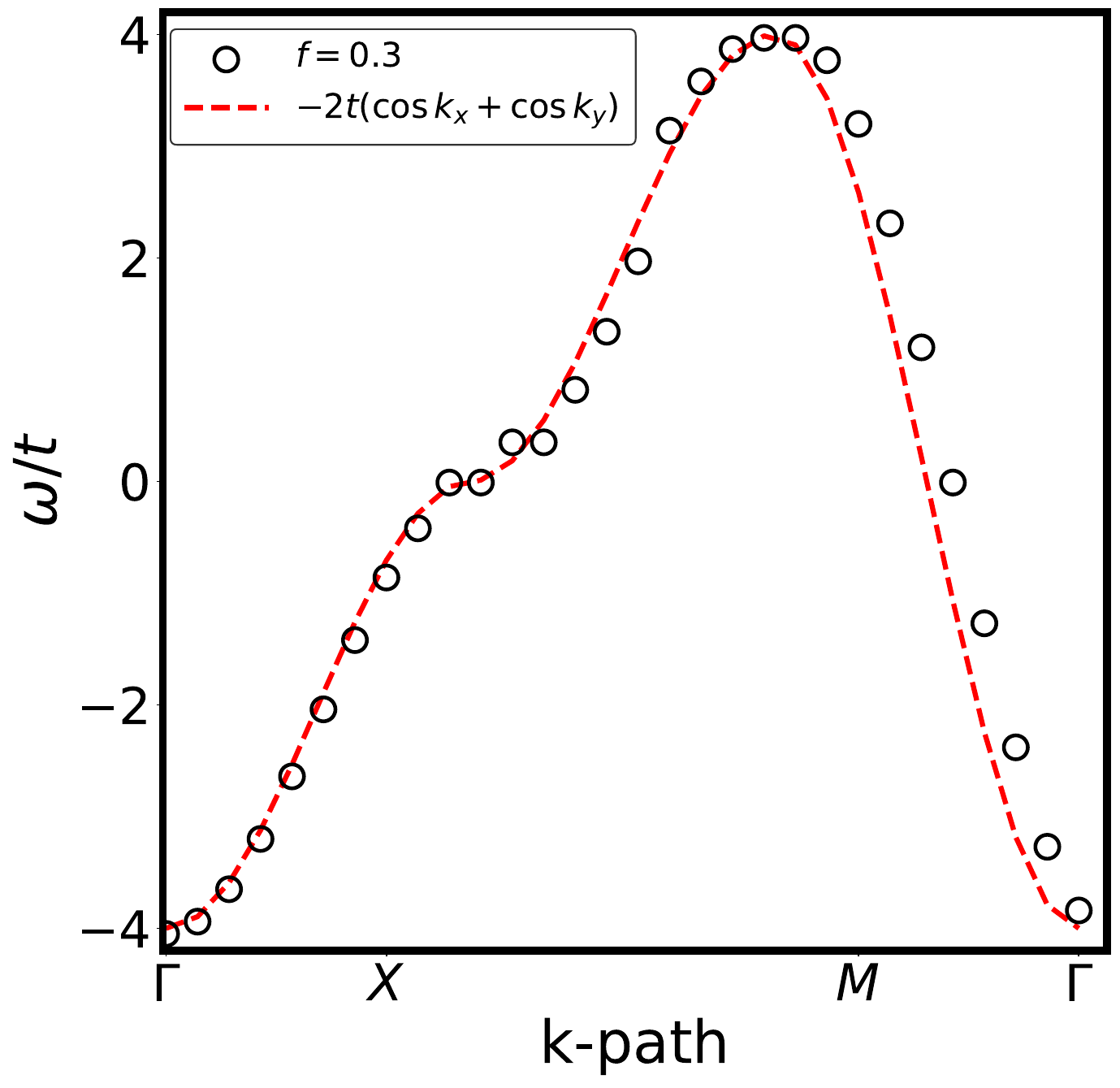}
\caption{Momentum--energy resolved spectral function
$A(\mathbf{k},\omega)$ along a high-symmetry $k$-path at $f=0.3$.
The spectral intensity follows the underlying bare dispersion
$\epsilon_{\mathbf{k}} = -2t(\cos k_x + \cos k_y)$, indicating
that the electronic band structure remains largely intact.
The symbol size represents the quasiparticle linewidth
(broadening).}
  \label{Fig:electronBand}
\end{figure}
% --------------------------------------------------------

The electronic spectral function is then defined as
\begin{equation}
A(\mathbf{k},\omega)
=
-\frac{1}{\pi}\,\mathrm{Im}\, G^R(\mathbf{k},\omega).
\end{equation}

This formulation allows momentum-resolved spectral information to be
reconstructed directly from the real-space eigenstates of the
inhomogeneous system. In practice, the spectral functions are averaged
over multiple realizations of the cluster geometry for each value of $f$.

\subsection{Momentum-resolved electronic spectral function}

We now examine the evolution of the momentum-resolved electronic spectral
function $A(\mathbf{k},\omega)$ as the fraction of interfacial sites $f$
is increased. Fig.\ref{Fig:electronspectra_momenta} 
shows $A(\mathbf{k},\omega)$ at three representative
momenta, $\mathbf{k}=(0,0)$, $(\pi,\pi)$ and $(\pi/2,\pi/2)$, for several
values of $f$. In the absence of interfacial inhomogeneity ($f=0$) the
spectral function exhibits sharp quasiparticle peaks corresponding to
well-defined electronic states.

As $f$ increases the most prominent effect is a systematic broadening
of the spectral peaks. This broadening reflects enhanced scattering of
electrons from the spatially varying lattice distortions 
near the interface. In our earlier transport study, these distortions
were shown to generate an effective static disorder potential that
controls the residual resistivity $\rho(0)$. The same scattering
mechanism manifests here as a finite quasiparticle lifetime, leading to
the observed broadening of $A(\mathbf{k},\omega)$ in
Fig.\ref{Fig:electronspectra_momenta}(a,b,c).

To quantify this effect we extract the linewidth $\Gamma_{\bf k}$ 
from
the spectral peaks.
The linewidth $\Gamma$ is extracted from
 $A(\mathbf{k},\omega)$ as the full width at half maximum 
 (FWHM) of the quasiparticle peak at each momentum.
Fig.\ref{Fig:electronspectra_momenta}(d) 
shows that $\Gamma$ increases
steadily with $f$ for all momenta considered. 
The growth
of $\Gamma$ directly reflects the increase of the electronic
scattering rate with increasing interfacial fraction. This
enhanced scattering is also consistent with the increase of the
residual resistivity $\rho(0)$ observed in related 
studies \cite{bose2024},
indicating that stronger scattering leads to both larger
linewidths in the spectral function and higher resistivity.
Note that that while the peaks broaden significantly
with increasing $f$, their positions shift only weakly - the
`band structure' effectively remains unrenormalised.

% --------------------------------------------------------
\begin{figure}[b]
\centering
\includegraphics[width=8.4cm,height=7.7cm]{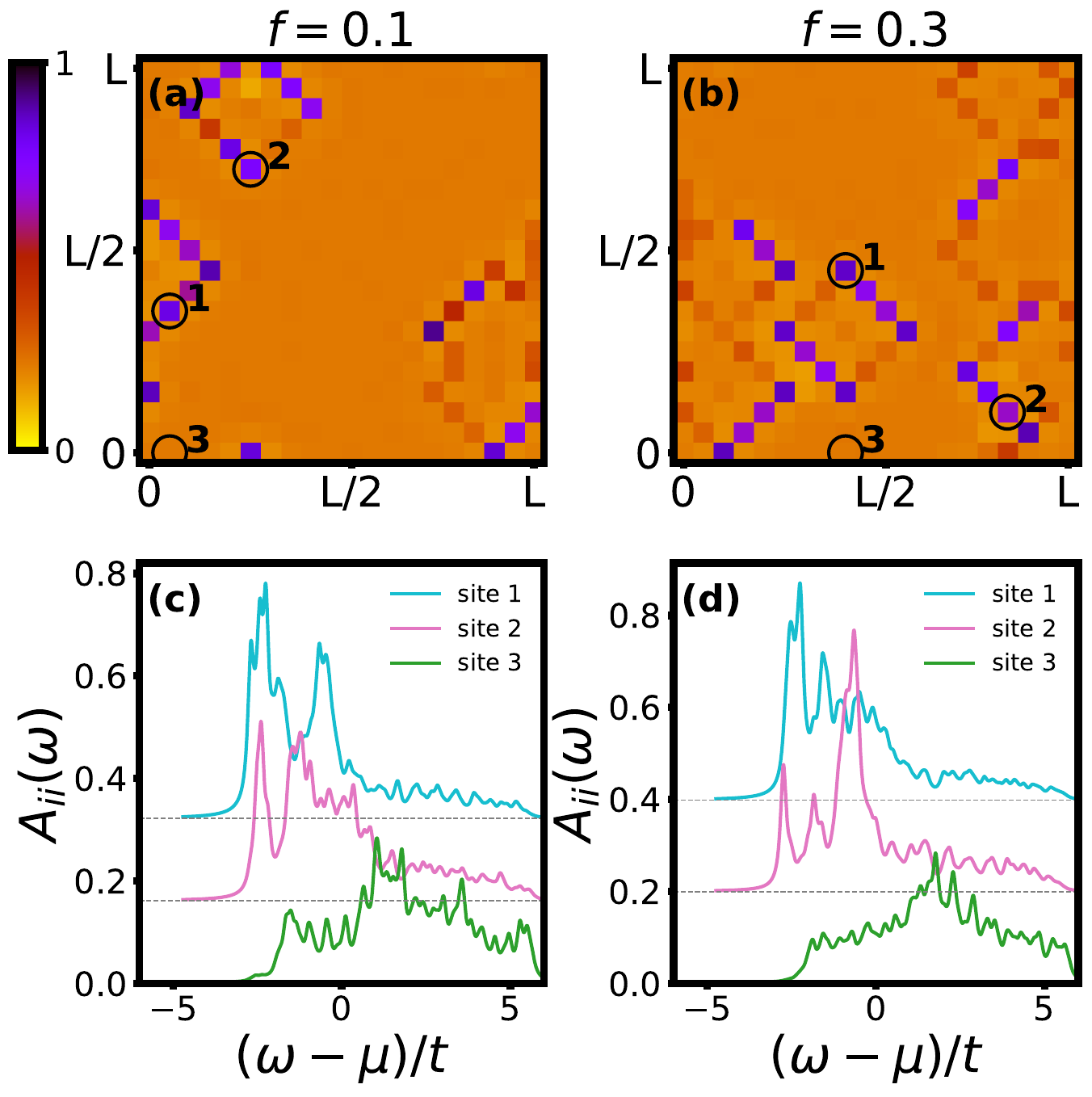}
\caption{Real-space electronic density and local spectral
 functions for (a,c) $f=0.1$ and (b,d) $f=0.3$. Panels (a,b)
 show spatial density maps, while (c,d) display the corresponding
 local spectral function $A_{ii}(\omega)$ at representative sites. 
 Site 3 corresponds to a bulk-like region with weak coupling, 
 site 1 lies on a strongly distorted interface with localized, 
 polaron-like states, and site 2 represents an intermediate 
 environment.  }
 \label{Fig:RealspaceelecronSpectra}
\end{figure}
% --------------------------------------------------------

 We plot the full
momentum-energy resolved spectral 
function $A(\mathbf{k},\omega)$ along
a high-symmetry momentum path.
Fig.~\ref{Fig:electronBand} shows the electronic spectral function at 
$f=0.3$ along the chosen $k$-path, overlaid with the bare band 
$\epsilon_{\mathbf{k}} = -2t(\cos k_x + \cos k_y)$. In the inhomogeneous case
the dispersion remains largely band like.
The width due to damping is roughly
the symbol size.

\subsection{Real-space electronic density and spectral function}

To understand the microscopic origin of the spectral broadening
discussed in the previous subsection, we examine the electronic
structure in real space. 
Fig.\ref{Fig:RealspaceelecronSpectra} shows the spatial distribution
of the electronic density together with the corresponding local
spectral function $A_{ii}(\omega)$ for two representative values
of the interfacial fraction, $f=0.1$ and $f=0.3$. The chemical
potential is fixed such that electron density($n$)=0.25.

In our model the bulk sites have weak electron--phonon coupling
$g_1$, while the interfacial sites have a much stronger coupling
$g_2$. The weak coupling $g_1$ leads to essentially uniform
low-density electronic states in the bulk regions. In contrast,
the strong coupling $g_2$ at the interface generates three
distinct types of electronic environments characterized by
low, intermediate, and high local electron density.

Three kinds of environments are illustrated by the 
the choice of representative
sites  in Fig.\ref{Fig:RealspaceelecronSpectra}.
 Site 3 lies in a bulk region and therefore
behaves similarly to a conventional two-dimensional tight-binding
system. Its local spectral function exhibits a broad distribution
of states typical of itinerant electrons. Because the calculation
is performed on a finite lattice, the spectral features appear
somewhat noisy, but the overall behavior remains consistent with
that expected for a two-dimensional tight-binding band.
The discussion below applies to both panels (c) and (d).

Site 1 has high electron density. 
The strong electron-phonon coupling produces
large local lattice distortions, leading to the formation of
polaron-like states. 
Consequently, the local spectral function shows a significant
transfer of spectral weight toward lower energies. This shift
reflects the polaron binding energy which lowers the local
electronic energy relative to the chemical potential.

Site 2 represents an intermediate-density environment located
near the interface. Although electrons are not fully localized
here, the nearby lattice distortions still influence the local
electronic states. As a result, the spectral weight is partially
shifted toward lower energies compared to the bulk site, reflecting
a tendency toward localization.

The coexistence of these three types of electronic environments
produces strong spatial fluctuations in the electronic potential.
These spatial
inhomogeneities act as effective scattering centers and provide
the microscopic origin of the linewidth broadening observed 
in the previous subsection.

\section{Phonon Spectra: Softening, Damping, 
and Spatial Inhomogeneity}

%Having examined the electronic spectral properties,
We now turn to the
lattice sector and analyze the phonon spectral function in the presence
of interfacial inhomogeneity. The phonon dynamics are obtained from the
electron–phonon interaction through the electronic polarization
function, which renormalizes the bare phonon propagator.

\subsection{Calculation of the phonon propagator}

The renormalized phonon propagator is constructed 
starting from the electronic
Green's functions obtained in the previous section. Using the
single-particle eigenstates of the electronic Hamiltonian, the
retarded real-space Green's function is given by Eq 9.
The electron–phonon interaction generates a
 polarization bubble which
renormalizes the phonon propagator. The real-space polarization
function \cite{freericks1998,esterlis2018} can be written as
\begin{eqnarray}
\Pi_{ij}(\omega) &=& g_i g_j \sum_{n,m}
\frac{
\Theta(\mu - E_n) - \Theta(\mu - E_m)
}{
\omega - (E_n - E_m) + i\eta
}
f_{ij}^{nm} 
\cr
f_{ij}^{nm} &=& 
\psi_i^{(n)} \psi_j^{(n)} \psi_i^{(m)} \psi_j^{(m)},
\end{eqnarray}
where $g_i$ and $g_j$ denote the local
 electron–phonon couplings and
$\Theta$ is the zero-temperature Fermi function.
Once the polarization function is obtained, the dressed phonon
propagator follows from a Dyson equation \cite{Heid},
\begin{equation}
[B(\omega)] =
[B^{(0)}(\omega)] +
[B^{(0)}(\omega)]\,
[\Pi(\omega)]\,
[B(\omega)].
\end{equation}
where $[B^{(0)}(\omega)]$, a matrix in {\it ij} space, 
is the bare phonon propagator.
This lead to 
\begin{equation}
\left[ B(\omega) \right] =
\left[ \mathbb{I} - B^{(0)}(\omega)\,\Pi(\omega) \right]^{-1}
\left[ B^{(0)}(\omega) \right].
\end{equation}
It is important to note that the renormalized propagator 
$B(\omega)$ defined above does not correspond to a 
quadratic phonon Hamiltonian with well-defined normal 
modes. The electronic polarization $\Pi_{ij}(\omega)$ is
 both frequency-dependent and complex, so the resulting 
 effective action describes a dissipative Gaussian theory
  for lattice fluctuations. As a consequence, phonon 
  excitations do not correspond to sharp eigenmodes even 
  in a fixed lattice configuration, and the spectral function 
  directly reflects both renormalization and intrinsic
   damping arising from electron--hole excitations.

% ---------------------

\begin{figure}[b]
\centering
\includegraphics[width=8.4cm,height=5cm]{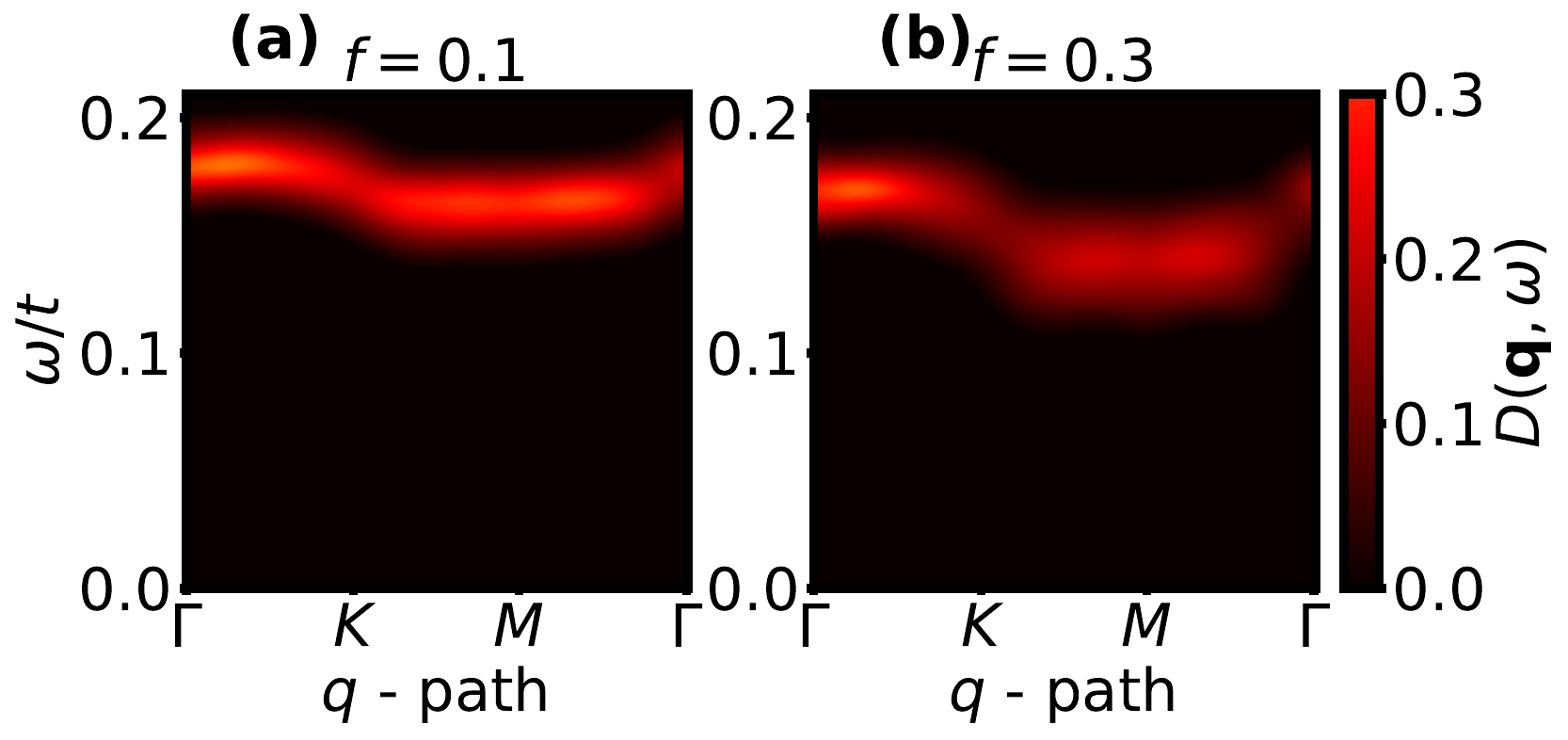}
\caption{
Momentum–energy resolved phonon spectral
 function $D({\bf q}, \omega)$ along a high-symmetry path
 for (a) $f = 0.1$ and (b) $f = 0.3$. With increasing
 interfacial fraction, the phonon response shows both
 a downward shift of spectral weight (softening) and
 a substantial broadening (damping). Since translational
 symmetry is broken, these spectra do not represent
 sharp dispersing normal modes but rather the projection
 of a spatially inhomogeneous, dissipative phonon system
 onto momentum space. The
 strongly coupled interfacial regions
 generate both local softening and enhanced
 decay into electronic excitations.
}
\label{Fig:phononDispersion}
\end{figure}

% ---------------------
% ---------------------

\begin{figure}[t]
\centering
\includegraphics[width=8.4cm,height=7.7cm]{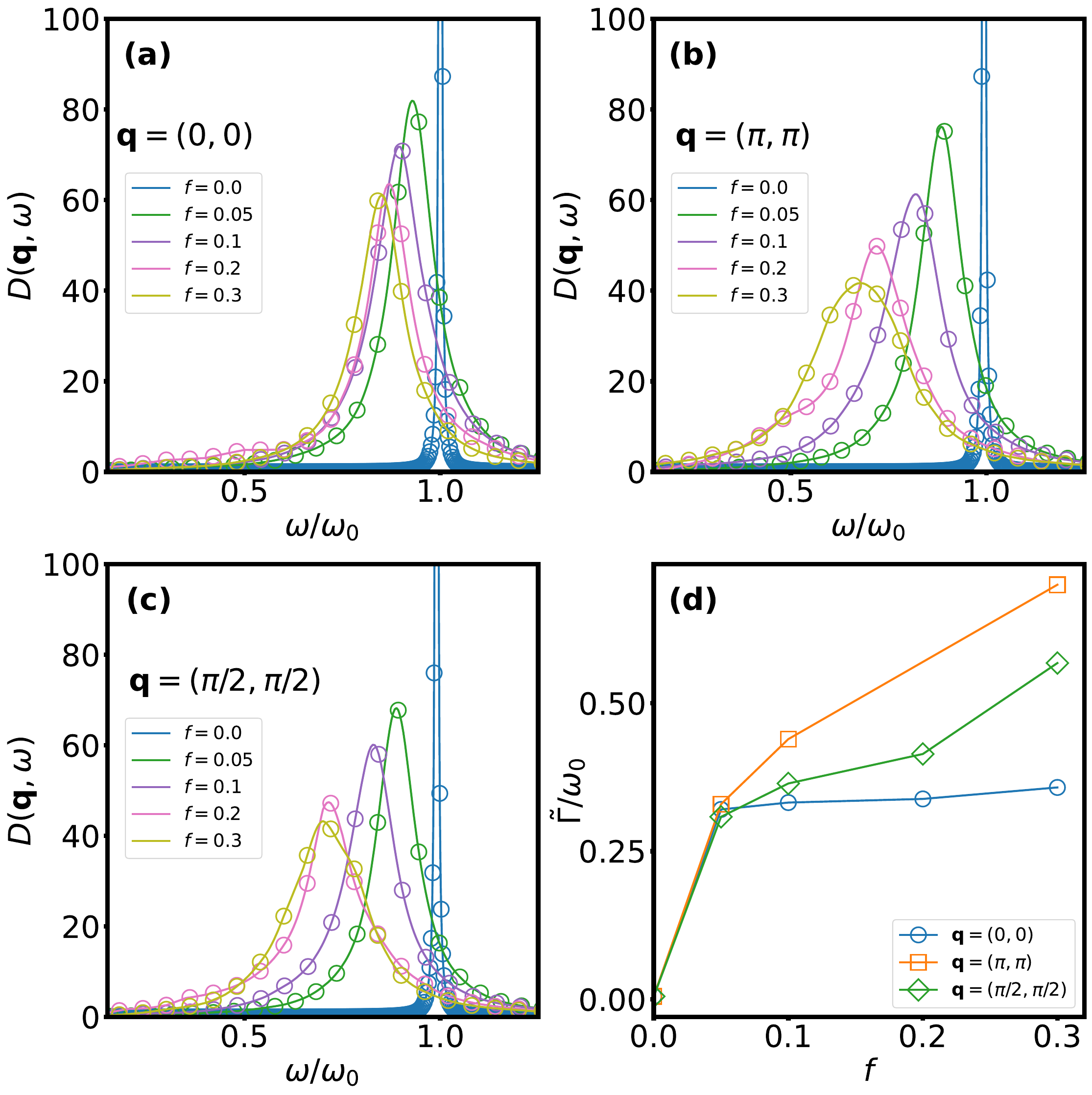}
\caption{
Line shapes of the phonon spectral function
 $D({\bf q}, \omega)$ at representative momenta (a) $q =
 (0, 0)$, (b) $(\pi, \pi)$, and (c) $(\pi/2, \pi/2)$ for increasing
  interfacial fraction $f$. Peaks shift to lower frequencies
  and broaden with increasing $f$, indicating simultaneous
  phonon softening and reduced phonon lifetimes. The
  softening reflects a reduction of the effective lattice
  stiffness arising from the static component of the electronic
   polarization, while the broadening originates from the
   frequency-dependent part of the polarization, which provides
   decay channels into low-energy electronic excitations.
   (d) Extracted linewidth $\Tilde{\Gamma}$ increases monotonically
   with $f$.}
 \label{Fig:PhononSpectra_momenta}
\end{figure}
 % ---------------------

Momentum-resolved phonon spectra are obtained by
Fourier transforming
the real-space propagator,
\begin{equation}
B(\mathbf{q},\omega) =
\sum_{ij}
e^{-i\mathbf{q}\cdot(\mathbf{r}_i-\mathbf{r}_j)}
B_{ij}(\omega).
\end{equation}
The phonon spectral function is then given by
\begin{equation}
D(\mathbf{q},\omega)
=
-\frac{1}{\pi}
\mathrm{Im}\,B(\mathbf{q},\omega).
\end{equation}

This procedure allows momentum-resolved phonon spectra to be
reconstructed directly from the real-space propagator even in the
presence of strong spatial inhomogeneity.

\subsection{Disorder-induced phonon renormalization and damping}

It is useful to distinguish two conceptually distinct
 contributions to the phonon renormalization. The static
  component, governed by $\Pi_{ij}(0)$, defines an 
  effective, spatially inhomogeneous stiffness matrix 
  $K^{\mathrm{eff}}_{ij} = K_{ij} - \Pi_{ij}(0)$, which
   captures the redistribution of phonon frequencies 
   and mode mixing due to broken translational
    invariance. In contrast, the frequency-dependent
     part of $\Pi_{ij}(\omega)$ generates an imaginary 
     component of the phonon self-energy, leading to 
     intrinsic damping through decay into electronic 
     excitations. The observed phonon spectrum 
     therefore reflects both static disorder effects 
     and dynamic electron-mediated processes.

We now examine how the phonon spectrum evolves with increasing
interfacial fraction $f$. Fig.\ref{Fig:phononDispersion} 
shows the momentum–energy
resolved phonon spectral function $D(\mathbf{q},\omega)$ along
a high-symmetry path in the Brillouin zone for two representative
values of $f$.

{For the homogeneous electron-phonon coupling, the 
harmonic phonons will have very small dispersion and 
damping}. For small disorder ($f=0.1$) the phonon 
spectrum remains relatively
sharp and follows a well-defined dispersion similar 
to the bare
phonon band.
 However, as the interfacial fraction increases
($f=0.3$) two important changes become apparent. First, the spectral
features broaden substantially, indicating enhanced phonon
damping. Second, the peak positions shift systematically toward
lower frequencies. This reduction of the phonon energy is a
signature of phonon renormalization and is commonly referred to
as \emph{phonon softening}
\cite{Bergmann,Takayama,esterlis2018breakdown}.

To quantify these effects we analyze line shapes 
of $D(\mathbf{q},\omega)$
at several representative momenta. Fig.\ref{Fig:PhononSpectra_momenta} 
shows the phonon spectral
function for $\mathbf{q}=(0,0)$, $(\pi,\pi)$ and $(\pi/2,\pi/2)$ for
several values of the interfacial fraction $f$.

As the interfacial fraction increases, the phonon peaks not only
broaden but also shift toward lower frequencies. The broadening
reflects a reduction of the phonon lifetime due to enhanced
scattering, while the shift of the peak position corresponds
to a renormalization of the phonon energy. The extracted linewidth
$\Tilde{\Gamma}_{\textbf{q}}$, shown in Fig.\ref{Fig:PhononSpectra_momenta}
(d), increases monotonically with $f$.
At large $f$, the phonon linewidth becomes comparable 
to the bare phonon energy scale, $\Tilde{\Gamma} \sim \omega_0$
 [see Figure~\ref{Fig:PhononSpectra_momenta}(d)].

The simultaneous presence of softening and damping originates
from the strong spatial inhomogeneity of the electron–phonon
interaction. In the bulk regions the coupling $g_1$ remains weak
and the phonon modes retain nearly their bare character.
However, at the interface the coupling $g_2$ is significantly
stronger, producing large local lattice distortions and strong
electron–phonon feedback. These strongly coupled regions act as
effective scattering centers for lattice vibrations. As phonons
propagate through the system they interact with electrons whose
states are already broadened by disorder, leading to a renormalized
phonon self-energy. This part is discussed in the subsection below.

\subsection{Local phonon spectra and lattice inhomogeneity}

\begin{figure}[t]
\centering
\includegraphics[width=8.4cm,height=7.7cm]{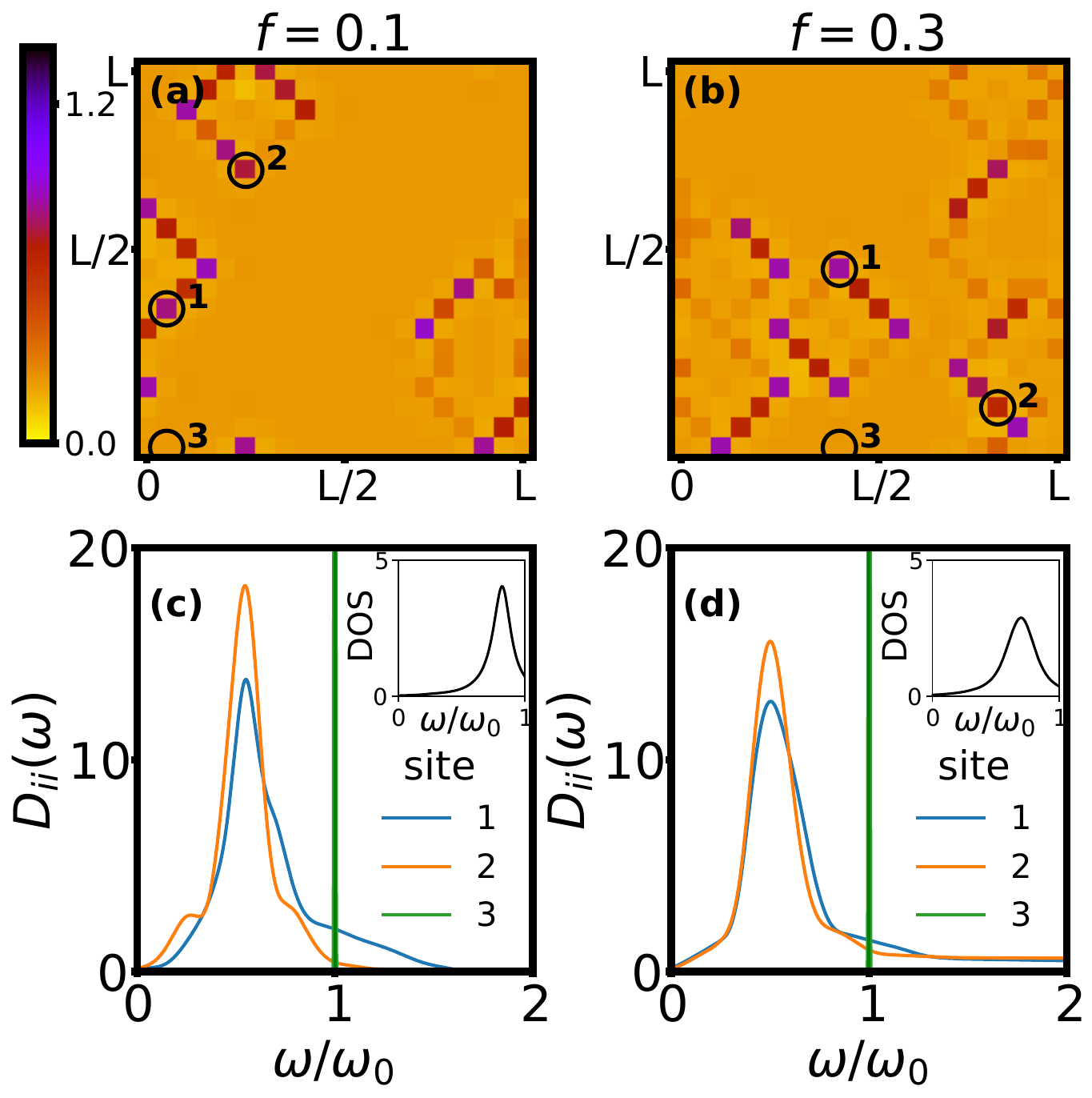}
\caption{
Real-space lattice distortions and local phonon
 spectra for (a,c) $f = 0.1$ and (b,d) $f = 0.3$. Panels (a,b)
  show spatial maps of lattice distortions, while (c,d) display 
  the corresponding local phonon spectral function 
  $D_{ii}(\omega)$ at representative sites. Bulk sites exhibit 
sharp
  peaks near the bare phonon frequency $\omega/\omega_0 
  \approx 1$, while interfacial sites show softened and 
  broadened peaks due to strong local electron–phonon coupling. 
The local softening can be interpreted in terms of a
 site-dependent effective stiffness, with interfacial sites 
 acting as regions of enhanced electronic susceptibility
  where lattice distortions can be accommodated at lower 
  energy cost. The broadening of 
  $D_{ii}(\omega)$ reflects coupling of local lattice
   fluctuations to a dense set of low-energy
    electronic excitations. 
}
\label{Fig:phononspectra_realspace}
\end{figure}

To understand the microscopic origin of the phonon renormalization
observed in momentum space, we examine the lattice distortions and
the corresponding local phonon spectral functions in real space.
Fig.\ref{Fig:phononspectra_realspace} shows 
representative lattice distortion maps together with
the local phonon spectral function $D_{ii}(\omega)$ for two values of
the interfacial fraction, $f=0.1$ and $f=0.3$. The frequency axis in
the spectral plots is normalized by the bare phonon frequency
$\omega_0 = 0.2$.

The lattice distortion maps reveal strong spatial variations that
arise from the inhomogeneous electron–phonon coupling. Bulk regions,
where the coupling remains weak ($g_1$), exhibit only small lattice
displacements. In contrast, interfacial regions with stronger
electron–phonon coupling ($g_2$) develop significantly larger
distortions due to enhanced local electron–lattice interactions.

% ---------------------
% ---------------------

These spatial variations in lattice distortions lead to distinct
local phonon environments. The three representative sites marked
in Fig.\ref{Fig:phononspectra_realspace} illustrate 
this behavior. Site 3 corresponds to a bulk
region where the lattice remains weakly distorted. 
The corresponding
local phonon spectral function exhibits a peak near
$\omega/\omega_0 \approx 1$, indicating that phonons at these
sites oscillate close to the bare phonon frequency. This behavior
is consistent with the weak coupling regime where lattice dynamics
remain largely unrenormalized.

In contrast, sites located near the interface experience stronger
lattice distortions. For these sites the local phonon spectral
function shows peaks that shift toward lower frequencies. 
Physically, the local
distortions modify the restoring force acting on the lattice,
thereby reducing the effective phonon frequency.

This local softening can be understood in 
terms of a site-dependent effective stiffness
\[
K_i^{\mathrm{eff}} = K -   \Pi_{ii}(0),
\]
where $\Pi_{ii}(0)$ is the local 
electronic susceptibility.
At zero temperature, $\Pi_{ii}(0)$ is directly related to 
 the local compressibility, and is therefore enhanced
  in regions where the electronic spectrum is dense 
  and responsive. In the present inhomogeneous 
  configurations, such conditions arise naturally at
  interfacial sites, where electronic states associated
  with bulk and strongly coupled regions overlap and
  exhibit small energy separations.

In addition to the frequency shift, the spectral features in
$D_{ii}(\omega)$ also broaden near the interface, indicating
enhanced damping of local phonon modes. The phonon density of
states (DOS), shown in the insets of 
Fig.\ref{Fig:phononspectra_realspace}, further illustrates
that spectral weight spreads toward lower frequencies as the
interfacial fraction increases. This redistribution of spectral
weight reflects the coexistence of phonons oscillating at nearly
bare frequencies in the bulk and softer modes emerging in the
distorted interfacial regions.

The presence of these spatially varying lattice environments
provides the real-space origin of the phonon renormalization
observed in momentum space. As phonons propagate through the
system they encounter regions with different local stiffness,
leading to both a downward shift of the dispersion and a
broadening of phonon spectral features.
Such modifications are expected to have a direct impact on 
  the effective electron-phonon interaction. To quantify 
  these effects  we now turn 
  to the Eliashberg spectral function.

\section{Eliashberg Spectral Function as a Measure
 of Inhomogeneous Coupling}

\label{sec:eliashberg}

Having analyzed the electronic and phonon spectra, 
we now examine how
the electron--phonon interaction manifests in a spectral sense in the
presence of spatial inhomogeneity. A commonly used quantity for this
purpose is the Eliashberg spectral function $\alpha^2F(\omega)$, which
characterizes the coupling between electronic states and phonons at
frequency $\omega$. The $\alpha^2F(\omega)$ can
 be extracted from the nonlinear current–voltage 
 characteristics of a point-contact spectroscope
  \cite{arind2,naidyuk2019point}. In the ballistic 
  regime, it is proportional to the derivative of point contact resistance.

In translationally invariant systems with well-defined quasiparticles,
$\alpha^2F(\omega)$ provides the basis for Migdal--Eliashberg theory and
allows a controlled description of superconductivity
 \cite{baggioli2020}. In the present
system, however, strong spatial inhomogeneity and substantial
quasiparticle damping complicate this interpretation. We therefore use
$\alpha^2F(\omega)$ primarily as a \emph{spectral measure of coupling},
rather than as an indicator of pairing interaction.

% ------------------------------------------------------
\begin{figure}[b]
\centering
\includegraphics[width=7.4cm,height=5.2cm]{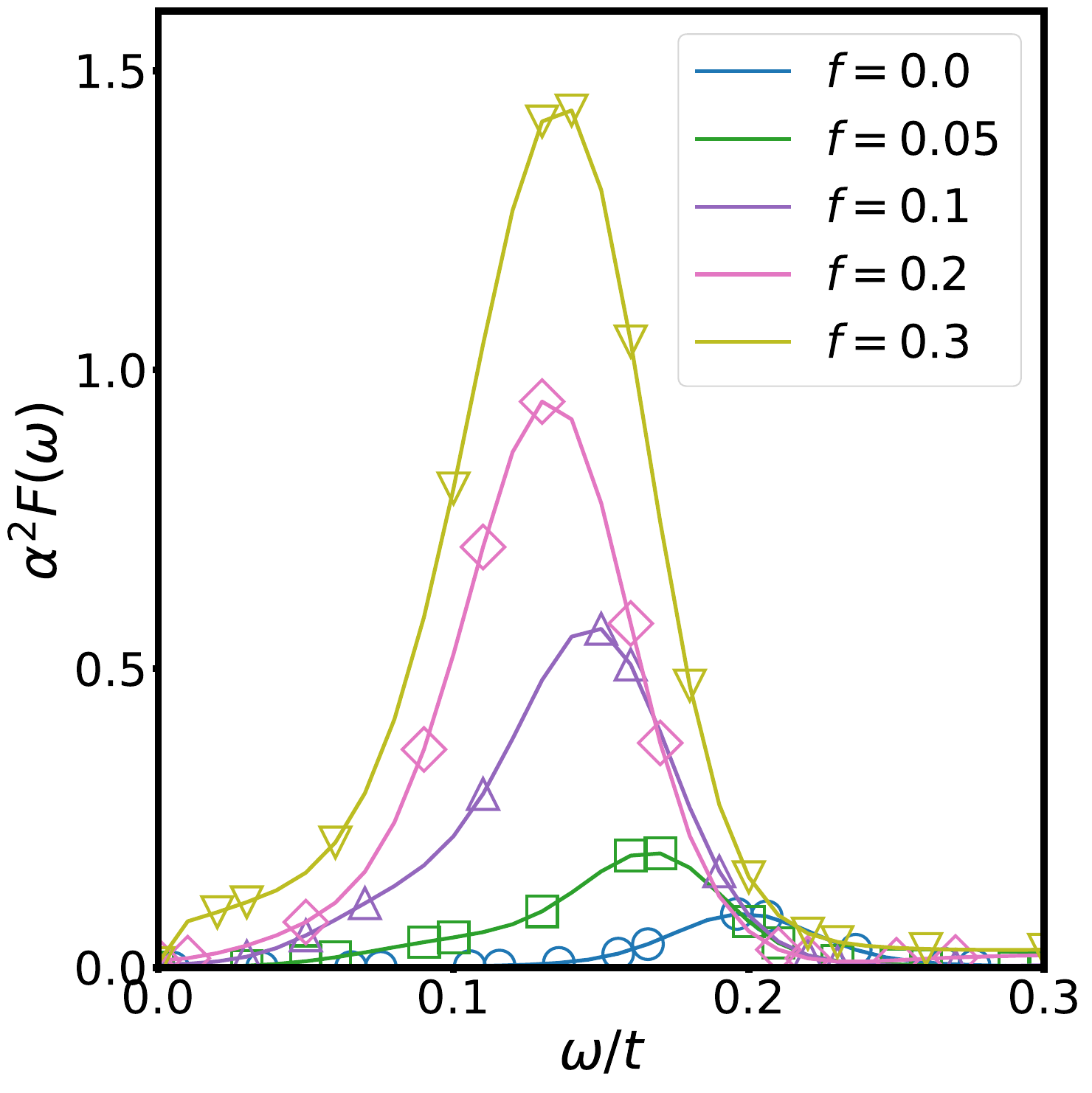}
\caption{Eliashberg spectral function $\alpha^2F(\omega)$ for 
increasing interfacial fraction $f$. With increasing $f$, spectral
 weight is progressively transferred toward lower frequencies, 
 reflecting enhanced coupling to softened phonon modes. 
 This redistribution of weight leads to a strong increase in the 
 effective coupling constant.}
 \label{Fig:Elaishberg}
\end{figure}
% ------------------------------------------------------
\begin{figure}[t]
\centering
\includegraphics[width=6.4cm,height=5.0cm]{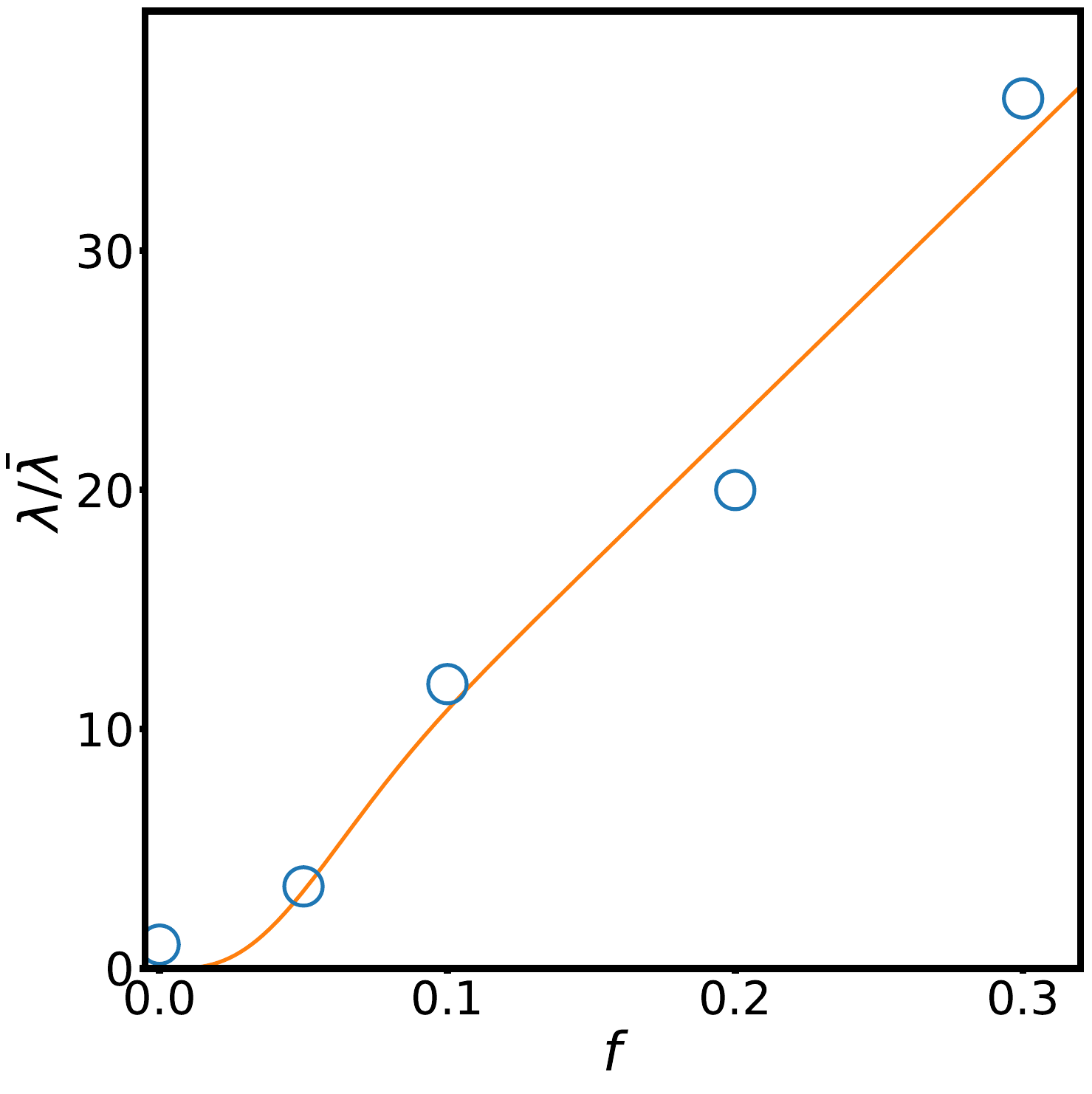}
\caption{
Effective electron-phonon coupling constant $\lambda$, 
normalized by its clean-limit value $\bar{\lambda} = 
\lambda(f=0)$, as a function of interfacial fraction $f$. 
The behavior is approximately linear at finite $f$, 
with a smooth suppression near $f \to 0$.
The increase of $\lambda$ with $f$ arises from 
enhanced low-frequency weight in $\alpha^2F(\omega)$, 
highlighting the strong influence of interfacial 
regions on the effective coupling.
}
    \label{Fig:effectivecoupling}
\end{figure}
% ------------------------------------------------------

\subsection{Calculation of the Eliashberg spectral function}

In translationally invariant systems the Eliashberg 
function\cite{Marsiglio,chubukov2020,Ummarino}
is expressed in momentum space as
\begin{equation}
\alpha^2F(\mathbf{k},\mathbf{k}',\omega)
=
N(\mu)\,
|g_{\mathbf{k},\mathbf{k}'}|^2
D(\mathbf{k}-\mathbf{k}',\omega),
\end{equation}
where $N(\mu)$ is the electronic density of states at the
Fermi level and $D(\mathbf{q},\omega)$ is the phonon spectral
function. However, in the present system translational symmetry
is broken by spatial inhomogeneity arising from the coexistence
of weakly coupled bulk regions and strongly coupled interfacial
sites. As a result, the phonon propagator is naturally computed
in real space.

Using the site-resolved phonon propagator $D_{ij}(\omega)$
obtained from the polarization function, the
momentum-resolved Eliashberg function can be written as
\begin{equation}
\alpha^2F(\mathbf{k},\mathbf{k}',\omega)
=
\sum_{i,j}
N(\mu)\, g_i g_j
D_{ij}(\omega)
e^{i(\mathbf{k}\cdot r_i - \mathbf{k}'\cdot r_j)} .
\end{equation}

This formulation naturally incorporates the spatial dependence of the
Holstein coupling and allows $\alpha^2F(\omega)$ to be evaluated
in the absence of translational symmetry. It should be emphasized,
however, that in such an inhomogeneous and strongly broadened system,
the connection between $\alpha^2F(\omega)$ and a well-defined
momentum-space pairing interaction is not straightforward.

To obtain the Fermi-surface averaged Eliashberg function
$\alpha^2F(\omega)$ we perform an average over electronic
states close to the Fermi level,
\begin{equation}
\alpha^2F(\omega)
=
\frac{1}{N(\mu)^2}
\sum_{\mathbf{k},\mathbf{k}'}
\alpha^2F(\mathbf{k},\mathbf{k}',\omega)
\delta(\epsilon_{\mathbf{k}}-\mu)
\delta(\epsilon_{\mathbf{k}'}-\mu).
\end{equation}

Finally, the effective coupling constant $\lambda$ is obtained
from the frequency integral of $\alpha^2F(\omega)$.

\subsection{Disorder evolution of the Eliashberg spectral function}

We first examine how the Eliashberg spectral function
$\alpha^2F(\omega)$ evolves as the degree of interfacial
inhomogeneity increases. Fig.\ref{Fig:Elaishberg} 
shows $\alpha^2F(\omega)$
for several values of the disorder parameter $f$.

In the clean limit ($f=0$), the Eliashberg function is
dominated by phonon modes near the bare phonon frequency.
As the fraction of interfacial sites increases, the
spectral weight of $\alpha^2F(\omega)$ progressively
shifts toward lower frequencies. This redistribution
of spectral weight directly reflects the phonon
renormalization discussed in the previous section,
where strong electron--phonon coupling at the
interfaces leads to local lattice distortions and
softened phonon modes.

The presence of spatial inhomogeneity therefore
introduces additional low-energy phonon excitations
that couple efficiently to electronic states near
the Fermi level. As a result, the Eliashberg spectral
function develops enhanced intensity at small
frequencies as $f$ increases.

\subsection{Enhancement of the effective coupling constant}

The effective coupling constant $\lambda$, defined as
\begin{equation}
\lambda = 2 \int_0^\infty d\omega \, \frac{\alpha^2F(\omega)}{\omega},
\end{equation}
is often used as a measure of the strength of electron--phonon
interactions. In conventional settings, $\lambda$ provides an estimate
of the pairing interaction entering Eliashberg theory.

In the present system, $\lambda$ should instead be viewed as a measure
of the redistribution of spectral weight toward low-frequency phonon
modes. Since the integrand is weighted by $1/\omega$, softened phonons
contribute disproportionately, leading to a rapid increase of $\lambda$
with increasing interfacial fraction [See Fig.\ref{Fig:effectivecoupling}].

This enhancement reflects the strong lattice distortions and phonon
softening at interfacial sites. At the same time, the electronic
spectra show significant quasiparticle broadening, indicating reduced
coherence. The coexistence of enhanced coupling and reduced electronic
coherence suggests that a large $\lambda$ in this system does not
necessarily translate into enhanced superconducting tendencies, but
rather signals strong interaction effects in an inhomogeneous
environment.

These results point to a separation between the magnitude of the
electron--phonon coupling, as measured by $\lambda$, and the emergence
of coherent many-body phenomena such as superconductivity. In
inhomogeneous systems, strong local coupling can coexist with strong
scattering and spatial fragmentation, potentially suppressing global
coherence. A proper treatment of pairing in this regime requires going
beyond conventional Eliashberg theory and is left for future work.

% -----------------

\section{Discussion: Interplay of Disorder and Interaction}

\label{sec:discussion}

The results presented above reveal a distinct regime of electron--phonon
physics arising from spatially inhomogeneous coupling, where strong
interactions are confined to interfacial regions embedded in a weakly
coupled metallic background. This situation differs qualitatively from
both homogeneous strong-coupling systems 
\cite{wellein1997polaron,hengsberger1999photoemission,
scalapino2018electron,li2010ground} and conventional disordered
metals \cite{al1979contribution,efetov1983supersymmetry}.

\subsection{Disorder versus interaction: nature of electronic scattering}

A central observation of this work is that increasing the fraction of
interfacial sites leads to a substantial broadening of the electronic
spectral function, while the underlying dispersion remains largely
unchanged. This indicates that the dominant effect of interfacial
inhomogeneity is to generate quasiparticle damping rather than band
renormalization. The electronic dispersion  can be 
measured via angle resolved photoemission spectroscopy 
(ARPES) \cite{ahn2004nonmetallic} and the local 
electronic spectra can be probed through scanning 
tunneling spectroscopy (STS) \cite {liu2006self} etc.

Microscopically, this behavior originates from lattice distortions
induced by strong electron--phonon coupling at the interface. These
distortions act as an effective, spatially fluctuating potential for
the electrons. In this sense, the system realizes a form of
\emph{self-generated disorder}, where the scattering landscape is not
externally imposed but arises dynamically from the interaction itself.

\subsection{Phonon renormalization in an inhomogeneous medium}

The lattice sector exhibits a complementary response. Phonon spectra
show both softening and significant broadening with increasing
interfacial fraction, indicating a simultaneous renormalization of
phonon energies and a reduction of phonon lifetimes.

Importantly, this renormalization is highly non-uniform in space. Bulk
regions retain nearly unrenormalized phonon modes, while interfacial
regions host strongly softened and damped excitations. The momentum-resolved
phonon spectra therefore represent an average over a distribution of
local environments. This coexistence of weakly and strongly renormalized
phonon modes is a key feature of the inhomogeneous system and has no
direct analogue in translationally invariant electron--phonon models. 
Phonon spectra can be experimentally probed using 
Raman and surface-enhanced Raman spectroscopy
 \cite{nandakumar2001raman, ramankutty2022molecule}.

From a theoretical perspective, the phonon spectrum can 
be viewed as arising from fluctuations in a self-generated, 
spatially inhomogeneous energy landscape. The electronic 
degrees of freedom produce both a static redistribution of
 local stiffness and a dynamic, frequency-dependent damping 
 kernel. The resulting phonon excitations therefore probe
  the structure of this landscape, revealing both soft 
  directions associated with locally enhanced susceptibility 
  and finite lifetimes arising from coupling to electronic 
  excitations.

\subsection{Enhancement of effective coupling and Eliashberg considerations}

The redistribution of phonon spectral weight toward low frequencies
leads to a pronounced enhancement of the effective electron--phonon
coupling constant $\lambda$. Since $\lambda$ is weighted by $1/\omega$,
softened phonon modes contribute disproportionately, resulting in a
rapid growth of the effective coupling with increasing interfacial
fraction.
This observation motivates a closer examination of 
how Eliashberg theory can be adapted to such inhomogeneous settings.

The presence of strong spatial
inhomogeneity suggest that the standard momentum-space formulation of
Eliashberg theory may not be directly applicable
\cite{Migdal1958,Eliashberg1960,alexandrov2001breakdown,
bauer2011quantitative,Grimaldi1995}. In this work we adopt
a real-space formulation of the Eliashberg function, which captures the
distribution of local couplings and phonon propagators, but a fully
self-consistent theory of pairing in such inhomogeneous 
systems remains
an open problem.

\subsection{Relation to effective-medium and homogeneous descriptions}

It is useful to contrast the present results with classical
effective-medium approaches 
\cite{Landauer1952,Kirkpatrick1973,Economou2006,Dobrosavljevic2012} 
In such descriptions, the system is
characterized by an average conductivity or coupling strength, and
microscopic spatial variations are integrated out. Our results show
that this averaging procedure misses essential physics. In particular,
the enhancement of low-frequency phonon modes and the associated
increase in $\lambda$ arise from strongly localized interfacial regions
and cannot be captured within a homogeneous description.

Similarly, the electronic spectra cannot be described in terms of a
uniform renormalized band structure with a single scattering rate.
Instead, the system exhibits a coexistence of itinerant and localized
electronic environments, leading to strongly momentum-dependent
broadening.

\subsection{Implications for nanohybrids and interface-driven phenomena}

The present results provide a microscopic framework for understanding
recent experiments on metallic nanohybrids, where large resistivity and
enhanced electron--phonon coupling have been reported. Our analysis
suggests that these effects originate from the interplay between strong
local coupling at interfaces and the extended nature of electronic
states.

More broadly, the results point to a regime in which interfaces act as
active elements that can strongly modify interaction effects without
significantly altering the underlying band structure. This opens the
possibility of engineering materials in which interaction strength and
scattering can be tuned independently through nanoscale structuring 
\cite{yu2015engineering}.

Overall, spatial inhomogeneity leads to substantial 
redistribution of spectral weight, enhanced damping, 
and modified quasiparticle characteristics across 
the system. These findings underscore the dominant 
role of interfacial regions in controlling the effective 
electron--phonon interaction. Such effects offer a 
promising pathway for engineering spectral 
response and pairing tendencies in nanostructured systems.

\subsection{Limitations and outlook}

Several aspects of the present work warrant further investigation. The
phonons have been treated at the harmonic level around equilibrium
configurations, and quantum fluctuations 
\cite{capone1997small} beyond this approximation may
play an important role at low temperatures. In addition, the present
analysis focuses on spectral properties; a fully self-consistent
treatment of transport \cite{allen2006,kumar2005} 
and superconducting correlations \cite{allen2000} in the presence
of spatially inhomogeneous coupling remains an open direction.

Finally, the question of how superconducting coherence emerges (or fails
to emerge) in a system with strongly enhanced but spatially localized
electron--phonon coupling is of particular interest, and will be
addressed in future work.

\section{Conclusion}
\label{sec:conclusion}

We have investigated the electronic and lattice spectral properties of
a metallic system with spatially inhomogeneous electron--phonon
coupling, motivated by nanohybrid structures with interface-driven
interactions. Using a real-space formulation of the Holstein model, we
have shown that strong coupling confined to interfacial regions leads
to a qualitatively distinct spectral regime.
For electrons the primary effect of this 
inhomogeneity is not a reconstruction of the
electronic band structure, but a reduction of quasiparticle
lifetimes arising from lattice distortions generated at the interface.
The phonon spectrum exhibits substantial softening
and damping, reflecting the coexistence of weakly and strongly
renormalized lattice environments.
These two effects combine to produce a significant redistribution of
the Eliashberg spectral function toward low frequencies, resulting in
a strong enhancement of the effective electron--phonon coupling
constant. The system therefore realizes a situation where 
interaction-induced
lattice distortions generate effective disorder, which in turn feeds
back on both electronic and phonon excitations.
Our results highlight the role of interfaces as active regions
that can strongly modify interaction effects without significantly
altering the underlying band structure. More generally, they point to
a route for engineering electronic and lattice properties through
controlled spatial inhomogeneity, beyond what is accessible in
homogeneous materials.
Future work will address the consequences of such inhomogeneous
coupling for superconducting correlations, particularly
the question of whether enhanced coupling can translate into coherent
pairing in the presence of strong spatial fluctuations.

\section{Acknowledgment}
The authors acknowledge use of the HPC clusters at 
HRI. S.Sarkar acknowledges support from  the 
Anusandhan National Research Foundation (ANRF) 
through a Grant No. PDF/2025/004884. 
% ----------------------------------------------------------------

\bibliographystyle{apsrev4-2}
\bibliography{v4}

\end{document}